\documentclass[10pt ]{revtex4}
\usepackage{hyperref}
\usepackage{amsmath,amsfonts}
\usepackage[dvips]{graphics}
\usepackage[dvips]{graphicx}
\usepackage{subfigure}
\usepackage{epsfig}
\usepackage{amsmath}
\usepackage{amsfonts}
\RequirePackage{color}

\begin{document}

\title{Thermodynamics and galactic clustering with a modified
gravitational potential}
  
  \author{Sudhaker Upadhyay}
   \email{sudhakerupadhyay@gmail.com}
  \affiliation{Centre for Theoretical Studies, 
Indian Institute of Technology Kharagpur,  Kharagpur-721302,  India}

\begin{abstract}
Based on thermodynamics, we study the galactic clustering of an expanding Universe by considering the logarithmic and volume (quantum) corrections to Newton's law  along with the repulsive effect of a
harmonic force induced by the cosmological constant  ($\Lambda$) in the formation of the large
scale structure of the Universe.
We derive  the $N$-body partition function  for 
extended-mass galaxies (galaxies with halos) analytically. For this
partition function, we compute the exact equations of states,
which exhibit the logarithmic, volume and cosmological constant corrections. In this setting,
  a modified  correlation (clustering) parameter (due to  these  corrections) 
  emerges naturally from the exact equations of state. We compute a  corrected grand canonical distribution function for this system. 
Furthermore, we obtain a deviation in
differential forms of the two-point correlation functions  for both the point-mass  
and   extended-mass cases. The  consequences of these deviations  on the correlation function's  power law are also discussed.
\end{abstract}
\maketitle
\textbf{Keywords}:  {Cosmology; Modified gravity; Galaxies cluster; Large scale structure of universe;

\hspace{1.9cm} Correlation function; Distribution function.}

\section{Overview and Motivation}

The characterization of galactic clusters on very large scales under the influence
of their mutual gravitational interaction  is a matter of vast interest.
The importance of such a process can be exaggerated as  the evolution and 
distribution of the galaxies throughout the Universe are the main manifestations of this.
The analysis of the correlation functions is  one of the standard ways to study the formation of the Universe. The observation tells us that the power law of two-point correlation function  scales as  $(\mbox{intergalactic distance})^{-1.6}$ to $(\mbox{intergalactic distance})^{-1.8}$ \cite{pee},
which has also been approved by $N$-body computer simulations \cite{ito}
and by the analytic gravitational quasiequilibrium
thermodynamics \cite{nas}.
The calculation of the power law of the correlation function is based on the assumption that 
the conversion of  the initial primordial matter  into the
observed many-body galaxies took place at the stage of evolution of the Universe  and these galaxies are coupled to the expansion of the Universe.
The theories of the   many-body (galaxies) distribution function have been developed mainly from a
thermodynamic point of view \cite{sas1,ahm02,nas,sas96,ahm10,ahm11}. Theses theories utilize
only the  first two
laws of thermodynamics   to
derive the exact equations of state  of the expanding Universe (a quasiequilibrium evolution).

The relation between thermodynamics and relativity
was originated in the  work of Bekenstein \cite{be}, Hawking \cite{ha}, and Unruh \cite{un}.
Later, Jacobson established an important connection between
  thermodynamics
and general relativity,  by which the Einstein equations themselves can be
viewed as  a thermodynamic equation
of state under a set of minimal assumptions involving  the equivalence principle and the identification of   the area of a causal horizon with   entropy \cite{a,b,c,c0,d}.
Recently, Verlinde  proposed a constructive   idea stating   that gravity is not a fundamental interaction
and can be interpreted as an entropic force  \cite{ver}. Although
this idea fails to provide a rigorous physical explanation \cite{ko,hu},    it  surely opens a new window into  understanding
  gravity from first principles.
For instance, the   modified Newton's law \cite{38},  Friedmann
equations at the apparent horizon of the Friedmann-Robertson-Walker Universe \cite{34,
35}, modified Friedmann equations \cite{36, 37},   Newtonian
gravity in loop quantum gravity \cite{39},   holographic dark energy \cite{40,41,42}, thermodynamics
of black holes \cite{43}, and  extension to Coulomb force \cite{44},  etc. 
support the entropic interpretation of gravity.

 Verlinde's approach  to get  Newton's law of gravity   relies
on the entropy-area relation  of   black holes in Einstein's gravity, i.e.,
$S=\frac{A}{4l^2_P}$, where $S$ is the entropy of the black hole, $A$ is the area of the horizon, and $l_P$ is the Planck length. 
In order to include the quantum corrections to the area law, some modifications 
are required to the area law \cite{sudp}.  In the literature,
the two well-studied  quantum
corrections to the area law are, namely, the logarithmic  correction and
power law correction. The logarithmic correction  appears due to the
thermal equilibrium fluctuations and the quantum fluctuations of   loop quantum gravity
\cite{45,46,47}. However, the power law correction appears due to the entanglement of quantum fields sitting near the horizon \cite{san,san1}.
Recently,    combined corrections   due to the logarithmic     and
  volume terms are  proposed for Newton's law of
gravitation as \cite{38}
\begin{eqnarray}
S=\frac{A}{4l^2_P}-a\log\left(\frac{A}{2l_P^2}\right)+b\left(\frac{A}{2l_P^2}\right)^{\frac{3}{2}},\label{ar}
\end{eqnarray}
where $a$ and $b$ are the constants of the order of unity or less. Here, the second term of the rhs corresponds to the logarithmic correction and 
the last term of the rhs corresponds to the volume correction.
Modesto  and  Randono   \cite{38} discussed   how  deviations from Newton's law caused by the logarithmic correction
have the same form as the lowest-order quantum effects of perturbative quantum gravity; 
however, the 
deviations caused by the volume correction follow the form of the modified Newtonian
gravity models explaining the anomalous galactic rotation curves.
In fact, on the very large (cosmological) scale,
 it is expected or otherwise speculated that
the cosmological constant, as a prime candidate for dark energy, is responsible for the 
expansion of the Universe through a repulsive force \cite{rind}.
If
the effect of the cosmological constant in the Newtonian limit of a metric is present in 
some phenomenon (such
as clustering of galaxies), then the same effect should also be present in the full general 
relativistic treatment of the same phenomenon.
Keeping the importance of the cosmological constant in mind,
 we want to employ the   cosmological-constant-induced (harmonic-oscillator-type) 
modification to   Newton's law, which leads to the cosmic repulsive force. Our motivation  here  is to study the effects of such corrections on the characterization of the clustering of galaxies
on very large scales.
The clustering of galaxies  within the framework of 
 modified Newton's gravity through the cosmological constant only has been studied very recently\cite{1}. 

To study the effects of the logarithmic, volume, and the cosmological constant corrections to  the 
clustering of  galaxies, we first derive the $N$-body partition function by  evaluating  configuration integrals recursively. From the resulting partition function,
we extract various thermodynamical (exact) equations of state. For instance, we compute   the
Helmholtz free energy,  entropy,  pressure, internal energy, and   chemical potential, which possess deviations from their original values due to  the logarithmic, volume, 
and cosmological constant corrections. Remarkably, a 
modified  correlation (clustering) parameter emerges naturally from the more exact  equations of state.  In the limit $a \rightarrow 0, b\rightarrow 0$, and $\Lambda\rightarrow 0$, the modified 
correlation  parameter  coincides with its original  value given in \cite{ahm02}. 
By assuming that the system is in a quasiequilibrium state  as described by the grand canonical ensemble, we derive the probability  distribution function.
The resulting distribution function depends on the modified clustering parameter.
Comparative analyses are made to see the effect of corrections on the
probability distribution function. In this regard, 
we find that the corrected distribution  function  first increases sharply with
the number of particles ($N$) and gets a maximum (peak) value for the particular
$N$. As long as $N$ increases further beyond that particular value,
the value of the distribution  function starts descending very fast and becomes slow 
later. The peak value of the corrected distribution  function 
decreases gradually with the increasing values of logarithmic and volume corrections.
Remarkably, the highly corrected distribution  function starts dominating
the lesser corrected distribution  function after a certain value of $N$.
Due to the cosmological constant term, the peak value of the corrected distribution  function 
decreases even further and  falls rather slowly.
We compute the differential form  of the
two-point correlation function for both the cases of point-mass and extended-mass   galaxies.
By solving the corresponding differential equation, we obtain a modified
structure of the two-point correlation function for both 
the point-mass and the extended-mass galaxies. It is shown that the  
 corrections also affect the power law of the correlation
function. Although there are  corrections  to the power law behavior, the
correlation function obeys 
the original result  (as in Refs. \cite{pee,ito,nas}) under certain approximations.

The paper is organized as follows. 
In Sec. II, we derive the $N$-body  partition function  
for the gravitationally interacting system with the corrected Newtonian dynamics using the
logarithmic, volume, and cosmological constant terms.
The   thermodynamical properties and distribution functions
for such a system are discussed in Sec.  III.  The differential form of the two-point correlation functions for both the point-mass and extended-mass galaxies are computed  in
Sec. IV. Within  this section, the effect of corrections on  the power law  behavior of
the two-point correlation functions is also discussed.
Finally,  the discussions and conclusions are made in the last section.

\section{ Interaction of galaxies through modified potential}
In this section, we consider a  modified Newton's law of gravitation
due to the first-order corrections and study their effects on the partition function.
   \subsection{A modified Newton's law of gravitation }
  It has been  stressed \cite{kau} that different quantum theories
of gravity may lead to different higher-order corrections to the area law of  
Bekenstein-Hawking entropy.
These corrections may display differences and, more interestingly,
 relations among quantizations. In  \cite{kau}, Kaul and Majumdar computed the
  lowest-order corrections to
the area law in a particular formulation  \cite{kau1} of a  quantum geometry 
program. They found that the leading correction is logarithmic, with
$\Delta S\sim \log (A/2l_P^2)$.   
On the other hand, in loop quantum gravity, the entropy  introduces a dependence
on the number of loops $L$ for the spin-network
state dual to a region of surface \cite{dd0}. In
the limit of a larger number of loops $L>>n$, where $n$ is the number of boundary edges, the 
entropy
behaves  as $S(L>>n) \sim  n \log L \sim n^{3/2}  \propto A^{3/2}$, where $L$ has exponential growth of the type $L\sim 2^{\sqrt{n}}$.
With these types of leading-order corrections to entropy, the 
expression of the (modified) area law results in  (\ref{ar}).
 
In fact,  
the 
Newtonian force ({\bf{F}}) in terms of entropy reads  ${\bf{F}}=-4l_P^2\frac{GM^2}{R^2}
\frac{\partial S}{\partial A}$. Therefore, corresponding to the logarithmic and the
volume   corrected entropy (\ref{ar}), 
   Newton's force law gives 
\begin{eqnarray}
{\bf{F}}=-\frac{GM^2}{R^2}\left[1-a\frac{l^2_P}{\pi R^2} +b12\sqrt{\pi}\frac{R}{l_P}\right].
\end{eqnarray}
 This leads to the following   corrected gravitational potential energy ($\Phi=-\int{\bf{F}} dR$)    \cite{38}:
\begin{equation}
\Phi =- {GM^2}\left[\frac{1}{R}-a\frac{l_P^2}{3\pi R^3} -b \frac{12\sqrt{\pi}}{l_P} 
\log
\left(\frac{R}{l}\right)  \right], \label{33}
\end{equation}
where   $l$ is an integration constant which signifies to 
(an unspecified) length parameter. 

However, at the cosmological scale, it is  speculated that
the cosmological constant $\Lambda$ is responsible  for the  expansion of the Universe
 through a repulsive force. 
For example, the Schwarzschild--de Sitter  spacetime in its static form is given
by the following line element \cite{dd}:
\begin{eqnarray}
ds^2=f(R)dt^2-\frac{1}{f(R)}dR^2 -R^2(d\theta^2 +\sin^2\theta d\phi^2),\nonumber\\
f(R)=\left(1-\frac{2GM}{R}-\frac{\Lambda R^2}{3}\right).\nonumber
\end{eqnarray}
Here we considered velocity of light $c=1$. From the line element, it is natural to include 
 an extra $\Lambda$-induced  harmonic-oscillator-type  potential   $-\frac{1}{6}\Lambda R$ \cite{rind} to (\ref{33}). Therefore, by incorporating the cosmological-constant-induced   modification     to  Newton's law, the  potential energy finally reads 
\begin{equation}
\Phi =- {GM^2}\left[\frac{1}{R}-a\frac{l_P^2}{3\pi R^3} -b \frac{12\sqrt{\pi}}{l_P} \log
\left(\frac{R}{l}\right) +\frac{1}{6}\frac{\Lambda R^2}{GM^2} \right].\label{phi}
\end{equation}
 In the next subsection, we will see the effect of these modifications 
on the many-body partition function.
   
   \subsection{The partition function}
 The statistical mechanics of an $N$-body system  is primarily based on the partition function.
Here we note that all our analyses  are based on the assumption that our gravitational  system has a statistically homogeneous distribution
over large regions,   which consists of an ensemble
of cells having the same volume $V$ and the same average density  $\bar \rho (N/V)$. 
In order to deal with the  galactic clustering from the statistical mechanics  
perspective, we first need to know the partition function ($Z_N(T,V)$) of the 
 gravitationally interacting system, which consists of $N$ particles of equal mass $M$,   momenta $p_{i}$ and average temperature $T$. 
This is generally given by \cite{ahm02} 
\begin{equation}
Z_N(T,V)=\frac{1}{\lambda^{3N}N!}\int d^{3N}{\bf p}\ d^{3N}{ R}\ \exp\left[-\frac{H}{T}
\right],\label{gene}
\end{equation}
where 
 $N!$ corresponds to the distinguishability of classical particles and $\lambda$ is a normalization constant  which results from the integration over momentum space. Here 
the $N$-body 
Hamiltonian has the following form: $H=\biggl[\sum_{i=1}^{N}\frac{p_{i}^2}{2M}+\Phi(r_1,r_2,...,r_N)\biggr]$.  
In general, the gravitational potential energy, $\Phi(r_{1}, r_{2}, \dots, r_{N})$,  
depends on the relative position vector of the $i${th}
and $j${th} particles (i.e., $R=|r_{i}-r_{j}|$) and, hence, describes the sum of the potential energies of all pairs. Therefore, $\Phi(r_{1}, r_{2}, \dots, r_{N})$ can be expressed as
\begin{equation}
 \Phi(r_{1}, r_{2}, \dots, r_{N})=\sum_{1\le i<j\le N}  \Phi_{ij}(R) = -T\sum_{1\le i<j\le N}
 \log (1+f_{ij}).
\end{equation} 
The two-point function $f_{ij}$  is introduced here to simplify the
  partition function  elegantly. The   $f_{ij}$ takes nonzero values only if there are  interactions present in the system. This function becomes negligibly small  for the system 
with asymptotically high temperature as well.

Upon integration over momentum space, the expression for the  partition function 
given in (\ref{gene}) reduces to
the following,
\begin{equation}
Z_N(T,V)=\frac{1}{N!}\left(\frac{2\pi MT}{\lambda^2}\right)^{3N/2}\Omega_N(T,V), \label{zn}
\end{equation}
where the  configurational integral $\Omega_{N}(T,V)$ has the following form:
\begin{equation}
\Omega_{N}(T,V)=\int....\int \prod_{1\le i<j\le N} (1+f_{ij})d^{3N}R. \label{q1}
\end{equation}
In this work, we will study the   clustering of galaxies interacting through
the modified  Newtonian  dynamics due to the   logarithmic, volume, and 
$\Lambda$-induced  corrections. 

It is evident from (\ref{phi}) that for the point-mass (galaxies') particles (i.e., $R=0$), the  potential energy  diverges. This leads to an ill-defined   Hamiltonian and, thus, the partition function. 
In order to remove this divergence, we consider  the extended nature of galaxies
 (galaxies with halos)  by introducing
a softening parameter $\epsilon$, which assures that the  galaxies are of finite  size.
The  softening parameter  takes a typical value $0.01\leq \epsilon\leq 0.05$ in units of the constant cell.   
 Thus, the effective potential
 energy modified by the logarithmic, volume, and $\Lambda$  terms, for the extended (real) mass galaxies in an expanding Universe, is given by
\begin{equation}
\Phi_{ij}(R)=- {GM^2}\left[\frac{1}{(R^2+\epsilon^2)^{1/2}}-a\frac{l_P^2}{3\pi (R^2+\epsilon^2)^{3/2}} -b \frac{6\sqrt{\pi}}{l_P} \log
\left(\frac{ R^2 }{l^2}\right)+\frac{1}{6}\frac{\Lambda R^2}{GM^2}   \right].\label{phii}
\end{equation}
 By definition,  we note  that  the systems  (that are still clustering)
  are not virialized on all scales, which implies that  the two-particle function $f_{ij}$  of the system  will be dominating up to linear order only. Thus, the two-particle function for the potential energy (\ref{phii}) is given by
\begin{eqnarray}
f_{ij}=\frac{GM^2}{T}\left[\frac{1}{(R^2+\epsilon^2)^{1/2}}-a\frac{l_P^2}{3\pi (R^2+\epsilon^2)^{3/2}} -b \frac{6\sqrt{\pi}}{l_P} \log
\left(\frac{ R^2  }{l^2}\right)  +\frac{1}{6}\frac{\Lambda R^2}{GM^2} \right].
\end{eqnarray}
Here  the higher-order terms are neglected.
Exploiting relation  (\ref{q1}),   the configuration integral  over a spherical volume of radius $R_{1}$ for $N=1$ (i.e., $f_{ij}=0)$ reads 
\begin{equation}
\Omega_{1}(T,V)=\int_0^{R_1} d^3 R =V.
\end{equation}
Now,  for $N=2$, the configuration integral $\Omega_{2}(T,V)$ (\ref{q1}) is calculated as
\begin{eqnarray}
\Omega_{2}(T,V)&=&4\pi V \int_{0}^{R_{1}}dR\ R^2\left[1+\frac{GM^2}{T}\left(\frac{1}{(R^2+\epsilon^2)^{1/2}}-a\frac{l_P^2}{3\pi (R^2+\epsilon^2)^{3/2}} \right.\right.\nonumber\\
&-&\left.\left. b \frac{6\sqrt{\pi}}{l_P} \log
\left(\frac{ R^2  }{l^2}\right)+\frac{1}{6}\frac{\Lambda R^2}{GM^2}   \right)\right].\nonumber
\end{eqnarray}
Here, in order  to convert the double integral into the single integral,  we consider the position of one particle (galaxy) to be fixed.  
 The expression of the configuration integral $\Omega_{2}(T,V)$ further simplifies to 
\begin{eqnarray}
\Omega_{2}(T,V)&=&V^2\left[1+  \frac{3}{2}  \frac{GM^2}{R_1T}\left( \sqrt{1+\frac{\epsilon^2}{R_1^2}} + \frac{\epsilon^2}{R_1^2} \log \frac{\epsilon/R_1}{\left[ 1+\sqrt{1+\frac{\epsilon^2}{R_1^2}}\right]} \right.\right. \nonumber\\
&+&\left.\left.  a\frac{2 l_P^2}{3\pi}\left(\frac{1}{R_1^2\sqrt{1+\frac{\epsilon ^2}{R_1^2}}}-\frac{1}{R_1^2\epsilon}+\frac{1}{R_1^2}\log \frac{{\epsilon/R_1}}{\left[ 1+\sqrt{1+\frac{\epsilon^2}{R_1^2}}\right]} \right)  \right.\right.\nonumber\\
&+& \left.\left.  b\frac{8\sqrt{\pi}}{l_P}\left( \frac{ R_1}{3}
  + {R_1} 
 \log \frac{l}{R_1}\right)\right)+ \frac{\Lambda R^2_1}{10T}\right].
\end{eqnarray}
In more compact form, this reads
\begin{equation}
\Omega_{2}(T,V)=V^2\biggl[1+ \frac{3}{2} (\alpha_1 +\alpha_2 +\beta_1 +\beta_2) \frac{ GM^2}{ R_{1}T}  \biggr],\label{a}
\end{equation}
which utilizes the following definitions:  
\begin{eqnarray}
 \alpha_1 \left( {\epsilon}\right)& =&  \sqrt{1+\frac{\epsilon^2}{R_1^2}} + \frac{\epsilon^2}
 {R_1^2} \log \frac{\epsilon/R_1}{\left[ 1+\sqrt{1+\frac{\epsilon^2}{R_1^2}}\right]},\ \ 
 \beta_1 = b\frac{8\sqrt{\pi}}{l_P}\left( \frac{ R_1}{3}
  + {R_1} 
 \log \frac{l}{R_1}\right).\nonumber  \\
 \alpha_2 \left( {\epsilon}\right)& =&a\frac{2 l_P^2}{3\pi}\left[\frac{1}{R_1^2\sqrt{1+
 \frac{\epsilon ^2}{R_1^2}}}-\frac{1}{R_1^2\epsilon} +\frac{1}{R_1^2}\log \frac{{\epsilon/
 R_1}}{\left[ 1+\sqrt{1+\frac{\epsilon^2}{R_1^2}}\right]} \right],\ \beta_2 =\frac{\Lambda R_1^3}{15 GM^2}.\label{para}
\end{eqnarray}
Since $\frac{GM^2}{R_1T}$ is a dimensionless quantity and remains invariant under   the scale transformations  $T\rightarrow\eta^{-1}T$ and $R_1\rightarrow \eta R_1$, then
according to the scaling property (see   Ref. \cite{lan} for details), we scale the quantity $\frac{GM^2}{R_1T}$ to $(\frac{GM^2}{R_1T})^3$  without the 
loss of generality.
Following this scaling, the expression (\ref{a}) reduces to
\begin{equation}
\Omega_{2}(T,V)=V^2\biggl[1+ \frac{3}{2} (\alpha_1 +\alpha_2 +\beta_1+\beta_2)  \left(\frac{ GM^2}{ R_{1}T}\right)^3\biggr].\label{ax}
\end{equation}
Since the radius of the cell  ($R_{1}$)  and  the average number
density per unit volume  [${\bar \rho} \sim (N/V)$]  are related through
 $R_{1}\sim ({\bar \rho})^{-1/3}$, then
\begin{equation}
  \frac{3}{2}\left(\frac{ GM^2}{ R_{1}T}\right)^3 \approx \frac{3}{2}\left(\frac{ GM^2}{  T}\right)^3{\bar \rho} := \omega.\label{sc}
\end{equation}
In terms of $\omega$, the configuration integral    (\ref{ax}) is expressed by
\begin{equation}
\Omega_{2}(T,V)=V^2\big[1+ (\alpha_1 +\alpha_2 +\beta_1+\beta_2) \omega\big].
\end{equation}
Following the  procedure mentioned above for $N$ particles recursively, we get
the most general  configuration integral as  follows: 
\begin{equation}
\Omega_{N}(T,V)=V^N\big[1+ (\alpha_1 +\alpha_2 +\beta_1+\beta_2)\omega\big]^{N-1}.\label{qn}
\end{equation}
Plugging the value of $\Omega_{N}(T,V)$ (\ref{qn})  simply into the general partition function (\ref{zn}),  we get
the following (explicit) expression for gravitational partition function:  
\begin{equation}
Z_N(T,V)=\frac{1}{N!}\left(\frac{2\pi MT}{\lambda^2}\right)^{3N/2}V^{N}\big[1+ (\alpha_1 +\alpha_2 +\beta_1+\beta_2)\omega\big]^{N-1}.\label{par}
\end{equation}
This is a  (canonical) partition function for a gravitational system of $N$ particles interacting through
the modified Newton's law. The corrections due to the
the logarithmic, volume,  and cosmological constant terms are inherent in the parameters $\alpha_2$, $\beta_1$, and $\beta_2$ respectively.
Once the expression of the partition function is known, it is   matter of calculating  the exact equations of state.  The expressions of the exact equations of state are important 
because these serve  as the primary ingredients  to evaluate all the thermodynamical properties  rigorously. 
\section{ Thermodynamics of galaxies under modified potential}
In this section, we first derive the various exact equations of state for the gravitating
system under the modified Newtonian dynamics. Later, we emphasize  the gravitational
 quasiequilibrium distribution function which originates from the  canonical  
partition function given in (\ref{par}).
\subsection{Exact equations of state}
In order to derive the  various equations of state, let us begin  
with the Helmholtz free energy.   The Helmholtz free energy is related to the general
 partition function  by $F=-T\log Z_{N}(T,V)$ (here Boltzmann's
constant is set to 1). 
Therefore, for the partition function given in (\ref{par}), the 
  Helmholtz free energy takes the following particular form:
\begin{equation}
F=-T\log\biggl[\frac{1}{N!}\left(\frac{2\pi MT}{\lambda^2}\right)^{3N/2}V^N\big[1+ (\alpha_1 +\alpha_2 +\beta_1+\beta_2) \omega\big]^{N-1}\biggr], \label{f}
\end{equation}
which further simplifies to
\begin{equation}
F=NT\log\left(\frac{N}{V}T^{-3/2}\right)-NT -NT\log\big[1+(\alpha_1 +\alpha_2 +\beta_1+\beta_2) 
\omega\big] -\frac{3}{2}NT\log\left(\frac{2\pi M}{\lambda^2}\right). \label{hel}
\end{equation}
Here, keeping the large value of $N$ in mind,  the   approximation $N-1\approx N$
is assumed.
Once the expression for the Helmholtz free energy is known, it is straightforward to
calculate various important thermodynamical entities, like   
pressure, entropy, and chemical potential, which are directly related to the Helmholtz free energy. For instance, the  entropy is related to the Helmholtz free energy in the following sense: $S= -\biggl(\frac{\partial F}{\partial T}\biggr)_{N,V}$.
 Therefore, corresponding to the Helmholtz free energy (\ref{hel}), the entropy reads
\begin{equation}
S=N\log\left(\frac{V}{N}T^{3/2}\right)+N\log\big[1+(\alpha_1 +\alpha_2 +\beta_1+\beta_2) \omega\big]-3N\frac{(\alpha_1 +\alpha_2 +\beta_1+\beta_2) \omega}{1+(\alpha_1 +\alpha_2 +\beta_1+\beta_2) \omega}+S_{0},\label{s}
\end{equation}
where the fiducial  entropy $S_{0}=\frac{5}{2}N+\frac{3}{2}N\log\left(\frac{2\pi M}{\lambda^2}\right)$. Now,  the specific entropy ($\frac{S}{N}$) is evident from the above as
follows:
\begin{equation}
\frac{S}{N}=\log\left(\frac{V}{N}T^{3/2}\right)-\log\left[1-\frac{(\alpha_1 +\alpha_2 +\beta_1+\beta_2) \omega}{1+(\alpha_1 +\alpha_2 +\beta_1+\beta_2) \omega}\right]-3\frac{(\alpha_1 +\alpha_2 +\beta_1+\beta_2) \omega}{1+(\alpha_1 +\alpha_2 +\beta_1+\beta_2) \omega}+s_0,
\end{equation}
where  $s_0=\frac{5}{2}+\frac{3}{2}\log\left(\frac{2\pi M}{\lambda^2}\right)$ is an arbitrary constant.
The entropy and   free energy  are related to the 
 internal energy $U$ through the relation $U =  F+TS$.
Therefore, exploiting the expressions of the free energy (\ref{f}) and   entropy (\ref{s}), the internal energy 
of the system is calculated as
\begin{equation}
U=\frac{3}{2}NT\left[1-2 \frac{(\alpha_1 +\alpha_2 +\beta_1+\beta_2)\omega}{1+(\alpha_1 +\alpha_2 +\beta_1+\beta_2)\omega}\right].\label{v}
\end{equation} 
The pressure and the Helmholtz free energy are related through the identity 
$P= -\left(\frac{\partial F}{\partial V}\right)_{N,T}$. This leads to the following  pressure equation of state
:
\begin{eqnarray}
  P=\frac{NT}{V}\left[1- \frac{(\alpha_1 +\alpha_2 +\beta_1+\beta_2)\omega}{1+(\alpha_1 +\alpha_2 +\beta_1+\beta_2)\omega}\right].\label{p} 
 \end{eqnarray}
The measure of the exchange of particles (chemical potential $\mu$)  for a given Helmholtz free energy (\ref{hel}) is calculated by the following relation $\mu = \biggl(\frac{\partial F}{\partial N}\biggr)_{V,T}$:
 \begin{eqnarray}
{\mu}&=&{T}\log \left(\frac{N}{V} T^{-3/2}\right)+{T}\log\left[1- \frac{(\alpha_1 +\alpha_2 +\beta_1+\beta_2)\omega}{1+(\alpha_1 +\alpha_2 +\beta_1+\beta_2)\omega}\right]-\frac{3}{2}{T}\log\left(\frac{2\pi M}{\lambda^2}\right)\nonumber\\
&-& \frac{(\alpha_1 +\alpha_2 +\beta_1+\beta_2)\omega}{1+(\alpha_1 +\alpha_2 +\beta_1+\beta_2)\omega}{T}.
 \end{eqnarray}
By drawing a comparison between the  exact equations of state obtained here 
 to their standard expressions (given in \cite{sas}),  we can classify the amount of corrections to 
 these equations of  the state.
 These expressions of Helmholtz free energy, entropy, free energy, pressure,
 and chemical potential 
 yield  the form of the modified clustering parameter ($\mathcal{B}_i$)  naturally.  
  This is given by
\begin{eqnarray}
 \mathcal{B}_i=\frac{(\alpha_1 +\alpha_2 +\beta_1+\beta_2)\omega}{1+(\alpha_1 +\alpha_2 +\beta_1+\beta_2)\omega}.
\label{b}
\end{eqnarray}
It is worth  evaluating the clustering (correlation) parameter,   as this provides the
 information regarding  the clustering of galaxies. The key feature  of the clustering   parameter  is as follows:
  in the typical limit of vanishing gravitational interaction, the clustering parameter
$\mathcal{B}_i$ tends to zero and, therefore, the system behaves as a perfect gas. However,
as long as $\mathcal{B}_i$ increases towards unity,  the system becomes more and more 
(strongly) bounded into clusters.   
The   corrected clustering parameter for the extended-mass structure (\ref{b}),
in terms of the original clustering parameter  (when there is no correction)
$b_\epsilon= {\alpha_1 \omega}/\left[{1+ \alpha_1 \omega}\right]$ \cite{ahm02},   is 
  given by
  \begin{eqnarray}
  \mathcal{B}_i  =\frac{b_\epsilon(1-\alpha_2 \omega -\beta_1\omega -\beta_2  \omega) +(\alpha_2 +\beta_1+\beta_2)\omega}{1 +(\alpha_2 +\beta_1+\beta_2)\omega -b_\epsilon (\alpha_2 \omega  +\beta_1  \omega+\beta_2 \omega)}.
  \end{eqnarray}
  Here we note that, as seen  from (\ref{para}),  the point-mass approximation (i.e.,
   $\epsilon =0$) is possible 
  for the volume and cosmological constant corrections  only (when $a=0$), but not for the logarithmic   case  (when $a\neq 0$) as the parameter $\alpha_2$ diverges in this approximation. 
\subsection{Gravitational quasiequilibrium distribution}
To study the distribution of voids, we assume that the system follows
a quasiequilibrium state, which is described by the equilibrium thermodynamics
at least as a first approximation \cite{sas1}. This requirement is met
by considering a grand canonical ensemble, where the number of galaxies and their mutual
gravitational energy vary among the members of the system.
 The   grand partition function  ($Z_{G}$) and the canonical partition function ($Z_{N}$)
 are related by \cite{sas1}
\begin{eqnarray}
Z_{G}(T,V,z)=\sum_{N=0}^{\infty}z^NZ_{N}(V,T), \label{g}
\end{eqnarray}
where $z$ is an arbitrary variable. 
The  grand partition function for the  gravitationally interacting system can be expressed
in terms of the thermodynamic variables as
\begin{equation}
\log Z_{G}=\frac{PV}{T}.
\end{equation}
Exploiting relation (\ref{p}), this further simplifies to
\begin{equation}
\log Z_{G}= \bar N(1- \mathcal{B}_i),\label{z} 
\end{equation}
where $\bar N$ refers to the average number of particles (which indicates   a grand canonical system).

The distribution function $F(N)$ for   finding $N$ particles in the energy state $U(N,V)$  (of the grand
canonical ensemble) is the sum over all of the energy states. It is given by
\begin{eqnarray}
F(N)=\frac{\sum_{i}e^{\frac{N\mu}{T}}e^{\frac{-U_i}{T}}}{Z_{G}(T,V,z)}=\frac{e^{\frac{N\mu}{T}}Z_{N}(V,T)}{Z_{G}(T,V,z)}.\label{fn}
\end{eqnarray}
Exploiting the values of $Z_N$ (\ref{par}) and $Z_G$ (\ref{z}),  
 the distribution function (\ref{fn}) for  finding $N$ extended-mass particles (galaxies) in the energy state $U(N,V)$  is computed as
\begin{equation}
F(N,\epsilon)=\frac{\bar{N}}{N!}\left(   \bar N( 1- \mathcal{B}_i ) + N \mathcal{B}_i  \right)^{N-1}\left( { 1- \mathcal{B}_i }\right)\exp{[-N \mathcal{B}_i -\bar N(1- \mathcal{B}_i)]}.
\end{equation}
Here, we observe that the structure of $F(N,\epsilon)$ is the same except for the
value of the clustering parameter $\mathcal{B}_i$.
Though  the  structure of the distribution function is similar to that of the unmodified Newtonian gravity case, the corrections due to the logarithmic, volume,
and cosmological constant terms are inherent in the clustering parameter
 $\mathcal{B}_i$.
Now, we would like to compare our results to those of the Ref. \cite{ahm02}. 
 The comparative study, as given in Figs. 1  and  2,  shows  that 
  $F(N)$ takes the maximum (peak) value for $N=\bar N$. As long as the values of   corrections  increase, the peak value of
$F(N)$ decreases. But as long as  the
  value of $N$ increases, after a particular $N$ the higher corrections of $F(N)$ dominate  their lower corrections. From these figures, it can be seen that
$F(N)$  decreases faster as the correction parameter  increases.
 \begin{figure}[htb]
 $%
\begin{array}{cc}
\epsfxsize=6cm \epsffile{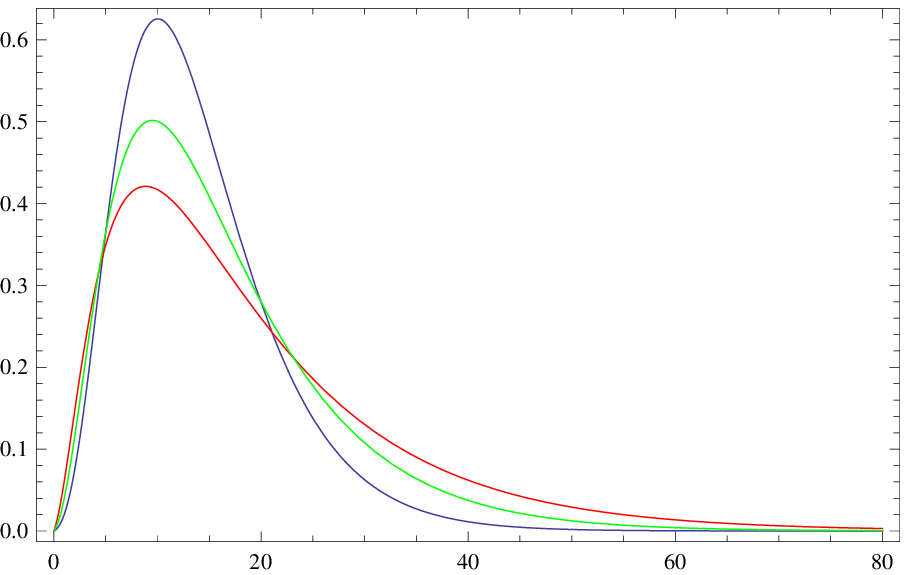} &\ \ \ \ \ \ \epsfxsize=6cm %
\epsffile{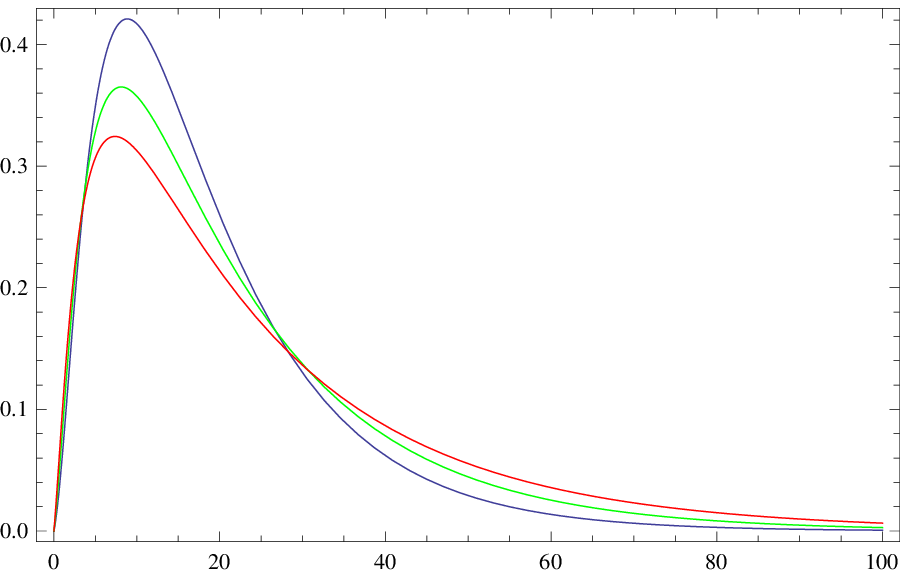} %
\end{array}
$%
 \caption{The distribution function $F(N)$ (perpendicular axis) versus $N$
 (horizontal axis) for an extended-mass structure for $b_\epsilon =0.5$ 
 and $\bar{N}=10$. Left: For   $\beta_2\omega=0$,  $(\alpha_2 +\beta_1)\omega =0$  corresponds to the violet line, $(\alpha_2 +\beta_1)\omega =0.5$ corresponds to 
the green line, 
 and $(\alpha_2 +\beta_1)\omega =1$  corresponds to the red line.
 Right: For   $\beta_2\omega=1$,  $(\alpha_2 +\beta_1)\omega =0$  corresponds to the violet line, $(\alpha_2 +\beta_1)\omega =0.5$ corresponds to 
the green line, 
 and $(\alpha_2 +\beta_1)\omega =1$  corresponds to the red line.
 }
 \end{figure}

\begin{figure}[htb]
 $%
\begin{array}{cc}
\epsfxsize=6cm \epsffile{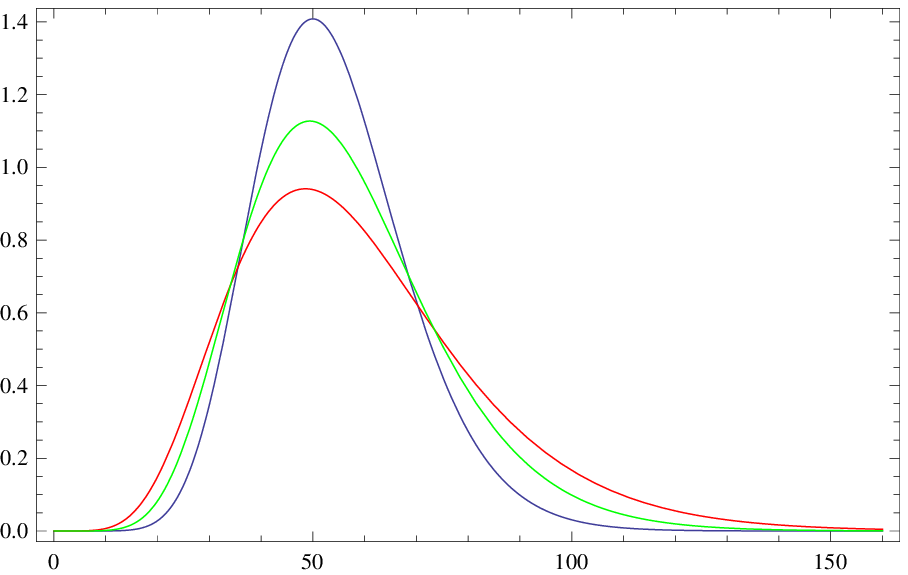} &\ \ \ \ \ \ \epsfxsize=6cm %
\epsffile{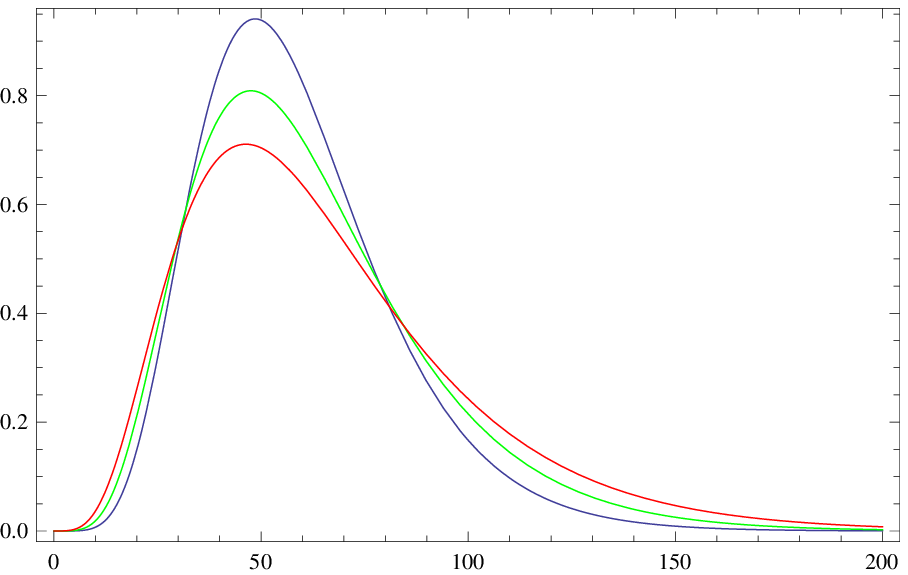} %
\end{array}
$%
 \caption{The distribution function(perpendicular axis) versus $N$
 (horizontal axis)  for an extended-mass structure for $b_\epsilon =0.5$  
 and $\bar{N}=50$. Left: For   $\beta_2\omega=0$,  $(\alpha_2 +\beta_1)\omega =0$  corresponds to the violet line, $(\alpha_2 +\beta_1)\omega =0.5$ corresponds to 
the green line, 
 and $(\alpha_2 +\beta_1)\omega =1$  corresponds to the red line.
Right: For   $\beta_2\omega=1$,  $(\alpha_2 +\beta_1)\omega =0$  corresponds to the violet line, $(\alpha_2 +\beta_1)\omega =0.5$ corresponds to  the
green line, 
 and $(\alpha_2 +\beta_1)\omega =1$  corresponds to the red line.}
 \end{figure}

  \section{The modified correlation functions}
  In this section, we compute the   two-point correlation functions under the modified Newton's law
for  the point-mass and extended-mass galaxies. This is done by  solving the
  differential forms  of the two-point correlation function. We analyze the power law 
  behavior of the corrected correlation functions also.  
  \subsection{Differential form for two-point correlation function}
Let us start the analysis by writing the internal energy for the homogeneous particles system   as 
 \cite{sas1} 
  \begin{eqnarray}
  U  &=&\frac{3}{2}NT +\frac{N\bar\rho}{2}\int_V \Phi(R) \xi_2({{\bar \rho}, R},T) dV,\label{int}
   \end{eqnarray}
 where  Boltzmann's constant is set to unit. Here, temperature $T$ represents   the random velocity; $\Phi(R)$ is a general interaction potential;  and $ \xi_2({{\bar \rho}, R},T)$
  is the two-point correlation function.
   The pressure equation of state for the gravitating system is given by \cite{sas96} 
   \begin{eqnarray}
  P  &=&\frac{NT}{V} -\frac{N\bar\rho}{6V} \int_V R\frac{d\Phi}{dR}\xi_2({{\bar \rho}, R},T) dV.\label{pot}
  \end{eqnarray}
Here the dynamical conditions for quasiequilibrium evolution are taken into account. 
   
 Corresponding to the specific   potential
 energy given in (\ref{phi}), the internal energy ($U_0$) and the pressure ($P_0$) equations of state  in the case of point-mass galaxies  read, respectively,
   \begin{eqnarray}
  U_0 &=&\frac{3}{2}NT   -  \frac{GM^2 N\bar\rho}{2}\int_V\left[\frac{1}{R}-a\frac{l_P^2}{3\pi R^3} -b \frac{12\sqrt{\pi}}{l_P} \log
 \frac{R}{l} +\frac{1}{6}\frac{\Lambda R^2}{GM^2} \right] \xi_2({{\bar \rho}, R},T) dV,\label{u00} 
 \end{eqnarray}
 and
 \begin{eqnarray}
  P_0 &=&\frac{NT}{V} -\frac{GM^2N\bar\rho}{6V} \int_V \left[\frac{1}{R} -a\frac{l_P^2}{\pi R^3}
  + b \frac{12\sqrt{\pi}}{l_P}-\frac{1}{3}\frac{\Lambda R^2}{GM^2}\right]\xi_2({{\bar \rho}, R},T) dV.\label{p0}
  \end{eqnarray}
Here we note that, for the grand canonical ensemble, the two-point correlation
function   $\xi_2$ depends on the variables ${{\bar \rho}, R}$, and $T$. So, one can write
 the differential  form of the  two-point correlation function as
  \begin{eqnarray}
  d\xi_2 = \frac{\partial\xi_2}{\partial\bar\rho} d\bar{\rho}+\frac{\partial\xi_2}{\partial
  T} dT+\frac{\partial\xi_2}{\partial R} dR.
  \end{eqnarray}
 This further leads to
   \begin{eqnarray}
  \frac{d\xi_2}{dV}  &=& -\frac{\bar{\rho}}{V}\frac{\partial\xi_2}{\partial\bar\rho}  + \frac{1}{4\pi R^2}\frac{\partial\xi_2}{\partial R},\label{diff}
  \end{eqnarray}
where $dT/dV=0$ is employed.
The Maxwell thermodynamic equation can be given in terms of   internal energy
$(U)$ and pressure $(P)$    by
  \begin{eqnarray}
  \left(\frac{\partial U}{\partial V} \right)_ {T,N} =T \left(\frac{\partial P}{\partial T} \right)_ {N,V} -P.\label{max}
  \end{eqnarray}
 Plugging the specific values of the internal energy (\ref{u00}) and pressure (\ref{p0}), obtained in the case of the point-mass galaxies, to the above Maxwell thermodynamic equation   and then differentiating the resulting equation with respect to $V$ leads to
  \begin{eqnarray}
  V  \frac{d\xi_2}{dV} =\left[ \frac{6\pi R^2l_PGM^2-6al_P^3GM^2 +72\pi^{3/2}bR^3GM^2 -2\pi \Lambda l_P R^5}{6\pi R^2l_PGM^2 -2al_P^3GM^2 -72\pi^{3/2}bR^3GM^2 \log \frac{R}{l} +\pi \Lambda l_P R^5}\right]T\frac{\partial\xi_2}{\partial T}.\label{se}
  \end{eqnarray}
 A first-order partial differential equation for the
   two-point correlation function in the case of a point-mass structure has the following form:
  \begin{eqnarray}
  3\bar{\rho}\frac{\partial\xi_2}{\partial \bar{\rho}}-R\frac{\partial\xi_2}{\partial R}
  +3\left[ \frac{6\pi R^2l_PGM^2-6al_P^3GM^2 +72\pi^{3/2}bR^3GM^2 -2\pi \Lambda l_P R^5}{6\pi R^2l_PGM^2 -2al_P^3GM^2 -72\pi^{3/2}bR^3GM^2 \log \frac{R}{l} +\pi \Lambda l_P R^5}\right]T\frac{\partial\xi_2}{\partial T} =0,
  \end{eqnarray}
where we have utilized expressions (\ref{diff}) and (\ref{se}).

By solving this first-order  differential equation, we get the explicit form for 
the two-point correlation
function as
\begin{eqnarray}
\xi_2({{\bar \rho}, R},T) =c(\bar{\rho})^{\frac{\chi}{3}}T^{\frac{\zeta}{3}} R^{\chi+3\zeta}\left(
2al_P^3GM^2 -6l_P\pi GM^2 R^2 +\pi\Lambda l_P R^5+72 b\pi^{3/2} GM^2 R^3 \log \frac{R}{l}
\right)^{- {\zeta}},\label{soln}
\end{eqnarray}
  where $c$ is an integration constant and  $\chi$ and  $\zeta$ are some arbitrary constant parameters.
  
 In order to get  the  differential equation  for 
the  two-point correlation in  the case of  an extended-mass structure,
we write the  expressions of internal energy $ U_{\mbox{ext}}$ and   pressure 
$P_{\mbox{ext}}$ as follows:
  \begin{eqnarray}
  U_{\mbox{ext}} &=&\frac{3}{2}NT - \frac{ {GM^2}N\bar\rho}{2}\int_V\left[\frac{1}{(R^2+\epsilon^2)^{1/2}}-a\frac{l_P^2}{3\pi (R^2+\epsilon^2)^{3/2}} -b \frac{12\sqrt{\pi}}{l_P} \log
 \frac{R}{l} +\frac{1}{6}\frac{\Lambda R^2}{GM^2}  \right] \xi_2  dV,\label{iex}\\
  P_{\mbox{ext}} &=&\frac{NT}{V} -\frac{{GM^2}N\bar\rho}{6V} \int_V \left[\frac{R^2}{(R^2+\epsilon^2)^{3/2}}-a\frac{l_P^2R^2}{\pi (R^2+\epsilon^2)^{5/2}} +b \frac{12\sqrt{\pi}}{l_P}  
 -\frac{1}{3}\frac{\Lambda R^2}{GM^2} \right]\xi_2  dV,
  \end{eqnarray}
which can be obtained simply  by exploiting  the relations (\ref{int}) and (\ref{pot}) together with potential energy (\ref{phii}).
  
Following the similar steps mentioned above for the case of point masses, we get a 
first-order partial differential equation for the two-point correlation
function for the extended-mass galaxies as follows:
  \begin{eqnarray}
 &&\frac{6\pi GM^2 R^2(R^2+\epsilon^2)l_P-6al_P^3GM^2R^2 +72\pi^{\frac{3}{2}}bGM^2(R^2+
 \epsilon^2)^{\frac{5}{2}}-2\pi\Lambda GM^2 l_P R^2(R^2+\epsilon^2)^{\frac{5}{2}}}{6\pi 
 GM^2(R^2+\epsilon^2)^2 l_P-2al_P^3GM^2(R^2+\epsilon^2) -72\pi^{\frac{3}{2}}bGM^2 (R^2+
 \epsilon^2)^{\frac{5}{2}}
   \log \frac{R}{l}+ \pi\Lambda GM^2 l_P R^2(R^2+\epsilon^2)^{\frac{5}{2}}}T\frac{\partial
   \xi_2}{\partial T}\nonumber\\
  &&  =\frac{R}{3}\frac{\partial\xi_2}{\partial R}-  \bar{\rho}\frac{\partial\xi_2}{\partial 
  \bar{\rho}}.\nonumber
  \end{eqnarray}
By solving this differential equation, we obtain an explicit expression
of the two-point correlation function in the case of extended-mass galaxies: 
 \begin{eqnarray}
\xi_2({{\bar \rho}, R},T) &=&C\bar{\rho}^{\frac{\chi}{3}}T^{\frac{\zeta}{3}} R^{\chi}(R^2+\epsilon^2)^{3\zeta/2}
\left[2 al_P^3GM^2 -6l_P\pi GM^2 (R^2+\epsilon^2) +\pi\Lambda GM^2 l_P^2 R^2(R^2+\epsilon^2)^{3/2} \right.\nonumber\\
&+&\left.72 b\pi^{3/2} GM^2(R^2+\epsilon^2)^{3/2} \log \frac{R}{l}
\right]^{- {\zeta}}.
\end{eqnarray}
Here, we notice that the above expression for the  two-point  correlation
function for an extended-mass structure in the limit $\epsilon\rightarrow 0$ reduces to 
the point-mass two-point  correlation
function (\ref{soln}).
  \subsection{Power law for correlation function}
Peebles's assumption that the two-point correlation function in a gravitational (galaxy) clustering obeys a power law \cite{pee} is in agreement with both the $N$-body computer simulations \cite{ito}   and analytic   gravitational quasiequilibrium thermodynamics \cite{nas}.
In order to  see the effects of the logarithmic, volume, and the cosmological constant deviated
  power law of the two-point correlation function, we write the correlation parameter as
   \begin{eqnarray}
  \mathcal{B}_i &=& \frac{{GM^2}{\bar \rho}}{6T}\int_V \left[\frac{1}{(R^2+\epsilon^2)^{1/2}}-a\frac{l_P^2}{3\pi (R^2+\epsilon^2)^{3/2}} -b \frac{6\sqrt{\pi}}{l_P} \log
\left(\frac{ R^2  }{l^2}\right)  +\frac{1}{6}\frac{\Lambda R^2}{GM^2}\right]\xi_2 ({{\bar \rho}, R},T) dV,\nonumber\\
&=&\frac{2\pi{GM^2}{\bar \rho}}{3T}\int_V \xi_2\left[\frac{R^2}{(R^2+\epsilon^2)^{1/2}}-a\frac{l_P^2R^2}{3\pi (R^2+\epsilon^2)^{3/2}} -b \frac{6\sqrt{\pi}}{l_P} R^2\log
\left(\frac{ R^2 }{l^2}\right)  +\frac{1}{6}\frac{\Lambda R^4}{GM^2} \right]   dR.
  \end{eqnarray}
This form of correlation parameter is obvious due to the expressions (\ref{v}) and (\ref{iex}).
 
By performing a differentiation with respect to $V$, this yields
  \begin{eqnarray}
 \frac{\bar \rho}{V} \frac{\partial \mathcal{B}_i }{\partial \bar \rho }  &=&   \frac{ \mathcal{B}_i }{ V }- \frac{{GM^2}{\bar \rho}}{6T}  \left[\frac{1}{(R^2+\epsilon^2)^{1/2}}-a\frac{l_P^2}{3\pi (R^2+\epsilon^2)^{3/2}} -b \frac{6\sqrt{\pi}}{l_P} \log
\left(\frac{ R^2  }{l^2}\right) \right.\nonumber\\
& +&\left. \frac{1}{6}\frac{\Lambda R^2}{GM^2} \right]\xi_2({{\bar \rho}, R},T),  
  \end{eqnarray}
where relation $
  \frac{\partial{\bar \rho}}{\partial V}=-\frac{\bar \rho}{V}
  $  is utilized. This further simplifies to  
  \begin{eqnarray}
  \frac{\bar \rho}{V} \left[ \frac{\partial \mathcal{B}_i }{\partial \bar \rho } -\frac{ \mathcal{B}_i }{\bar{\rho}}\right]
   =  - \frac{{GM^2}{\bar \rho}}{6T}  \left[\frac{1}{(R^2+\epsilon^2)^{1/2}}-a\frac{l_P^2}{3\pi (R^2+\epsilon^2)^{3/2}} -b \frac{6\sqrt{\pi}}{l_P} \log
\left(\frac{ R^2 }{l^2}\right) +\frac{1}{6}\frac{\Lambda R^2}{GM^2}  \right]\xi_2. \label{gf}  
  \end{eqnarray}
The relation (\ref{b}) yields 
  \begin{eqnarray}
  \frac{\partial{\cal B}_i}{\partial{\bar \rho}}  -\frac{ \mathcal{B}_i }{\bar{\rho}}= \frac{ -{\cal B}^2_i }{{\bar \rho}}. \label{bu}
  \end{eqnarray}
Now,  by substituting the value of (\ref{bu}) to   Eq. (\ref{gf}), we obtain
   \begin{eqnarray}
  \frac{{\cal B}^2_i}{V}
   =  \frac{{GM^2}{\bar \rho}}{6T}  \left[\frac{1}{(R^2+\epsilon^2)^{1/2}}-a\frac{l_P^2}{3\pi (R^2+\epsilon^2)^{3/2}} -b \frac{6\sqrt{\pi}}{l_P} \log
\left(\frac{ R^2   }{l^2}\right) +\frac{1}{6}\frac{\Lambda R^2}{GM^2}  \right]\xi_2.   
  \end{eqnarray}
Consequently, we get the power law behavior of the two-point correlation function for
 the extended-mass galaxies  as follows:
  \begin{eqnarray}
 \xi_2  = \frac{{9T} {\cal B}^2_i/ {GM^2}{\bar \rho} }{  \left[\frac{2\pi R^3}{(R^2+\epsilon^2)^{1/2}}-a\frac{2 R^3l_P^2}{3  (R^2+\epsilon^2)^{3/2}} -24\pi^{3/2}b \frac{ R^3 }{l_P} \log
 \frac{ R}{l}  +\frac{\pi \Lambda R^5}{3GM^2} \right]}.
  \end{eqnarray}
In the limit of point masses, this reduces to
   \begin{eqnarray}
 \xi_2  = \frac{{9T} {\cal B}^2_i/ {GM^2}{\bar \rho} }{  \left[ {2\pi R^2} 
 -\frac{2}{3}al_P^2 -24\pi^{3/2}b \frac{ R^3 }{l_P} \log
 \frac{ R}{l} +\frac{\pi \Lambda R^5}{3GM^2}  \right]},\label{xa}
  \end{eqnarray} 
  which, eventually, characterizes  the
  modification to the power law equation given in \cite{nas}. 
These modifications are due to the logarithmic, volume, and $\Lambda$ terms. 
 From observation,
the correlation function has a simple power law \cite{pee},
\begin{eqnarray}
 \xi_2= (R_0/R)^{\gamma},\label{xxx}
\end{eqnarray}
where $\gamma\sim 1.77$ and $R_0\sim 5.4 h^{-1}Mpc$. 
 This apparent simplicity has led many investigators
to describe theoretical results and numerical simulations. However, 
$\xi_2$ possesses very limited information. Later
observational analyses give a considerable range for $\gamma$ and $R_0$, 
which suggests that $\xi_2$ may not have a simple power law as given in (\ref{xxx}); rather it requires
the higher-order corrections to the correlation function. In this regard,  the  
corrected correlation function having expression  (\ref{xa}) may be of interest.
The  quantitative discussions, like the particular scale on which
these modifications will be important, are the subject of further investigation.
Remarkably, we note here that the  power law still behaves  as $\xi_2 \sim R^{-2}$  at leading order in the $l_P\sim R$ approximation.

\section{Discussion and Conclusions}
In order to study  the clustering of galaxies in an expanding Universe under the 
  modified  Newton's law, we have considered a  logarithmic, volume, and the cosmological constant modified Newtonian potential. 
Corresponding to this modified Newtonian potential, we have calculated an explicit expression for the (canonical) partition function, which describes $N$  extended-mass galaxies, with the help of configuration integrals.
Our whole calculation  is based on the assumption that the system (ensemble),  which is made
of cells of the same 
volume $V$ and average density $\bar \rho$, has statistically homogeneous distribution over large regions. With the help of the resulting partition function,
we have derived various   thermodynamical equations of state, namely, the Helmholtz free energy,
  internal energy,  pressure,   entropy, and   chemical potential equations of state.
As a consequence of these exact equations of state, a corrected 
correlation  (clustering) parameter emerges naturally for the clusters of the galaxies with halos. 
Here, we have found that the point-mass approximation is possible only for the volume 
correction  but not for the logarithmic and $\Lambda$-induced corrections as the   parameter  
$\alpha_2$ only diverges in the point-mass approximation.
Nevertheless, this is not a serious threat as all the (real) galaxies are of finite size  as 
described by the softening parameter $\epsilon$.
Moreover, we have derived a modified version of the distribution function $F(N)$ (probability of finding $N$ 
gravitating bodies in volume $V$) for the gravitating system which follows the
quasiequilibrium and, thus,  resembles  the grand canonical ensemble.
We note that the corrections  on the distribution function are inherent in the
 clustering parameter. The behavior of these corrections on 
the distribution function $F(N)$ is discussed through the plot (see, e.g., Figs. 1 and 2 above), where we observe  that the peak (maximum) values of $F(N)$ (at $N=\bar N$) decrease  
as the  correction dominates. As long as $N$ increases, after a certain value of $N$ the 
higher value of the correction  starts dominating the smaller one because the values of $F(N)$ decrease faster for smaller  corrections. The presence of the cosmological constant, 
which is responsible for the   expansion of the Universe through a repulsive force,
reduces  the peak value (at $N=\bar{N}$)  of the corrected  distribution function even further, but makes the  descent of the distribution function slower.
 
 We have also derived the  two-point correlation function ($\xi_2$)  for the gravitating system under this modified Newton's law  for the cases of both 
 point masses and extended masses. In this regard, we have first calculated the differential
 form of the modified two-point correlation function.
The solution of the  differential equation leads to the exact form of the two-point correlation function, where deviations from the original value are evident. In the limit $\epsilon\rightarrow 0$, the extended-mass two-point correlation function coincides to that of 
the point-mass case. The effect of corrections on the
 power law of the correlation function is also discussed and it has been found that it behaves as $\xi_2\sim R^{-2}$ at leading order in a certain approximation, which is 
consistent with the result obtained in Ref. \cite{nas}.

\end{document}